\documentclass[aps,prl,twocolumn,superscriptaddress,showpacs,floatfix]{revtex4}

\usepackage{graphicx}
\usepackage{epstopdf}
\usepackage{amssymb}
\usepackage{amsmath}
\usepackage{color}
\usepackage{bm}

\begin{document}

\title{Search for Neutrinoless Double-Beta Decay in $^{\bm{136}}$Xe with EXO-200}



\author{M.~Auger}
 \affiliation{LHEP, Albert Einstein Center, University of Bern, Bern, Switzerland}
\author{D.J.~Auty}
 \affiliation{Department of Physics and Astronomy, University of Alabama, Tuscaloosa AL, USA}
\author{P.S.~Barbeau}
\altaffiliation{Corresponding author: psbarbea@stanford.edu}
 \affiliation{Physics Department, Stanford University, Stanford CA, USA}
\author{E.~Beauchamp}
 \affiliation{Department of Physics, Laurentian University, Sudbury ON, Canada}
\author{V.~Belov}
 \affiliation{Institute for Theoretical and Experimental Physics, Moscow, Russia}
\author{C.~Benitez-Medina}
 \affiliation{Physics Department, Colorado State University, Fort Collins CO, USA}
\author{M.~Breidenbach}
 \affiliation{SLAC National Accelerator Laboratory, Stanford CA, USA}
\author{T.~Brunner}
  \affiliation{Physics Department, Stanford University, Stanford CA, USA}
\author{A.~Burenkov}
 \affiliation{Institute for Theoretical and Experimental Physics, Moscow, Russia}
\author{B.~Cleveland}
 \altaffiliation{Also SNOLAB, Sudbury ON, Canada}
 \affiliation{Department of Physics, Laurentian University, Sudbury ON, Canada}
\author{S.~Cook}
 \affiliation{Physics Department, Colorado State University, Fort Collins CO, USA}
\author{T.~Daniels}
 \affiliation{Physics Department, University of Massachusetts, Amherst MA, USA}
\author{M.~Danilov}
 \affiliation{Institute for Theoretical and Experimental Physics, Moscow, Russia}
\author{C.G.~Davis}
 \affiliation{Physics Department, University of Maryland, College Park MD, USA}
 \author{S.~Delaquis}
 \affiliation{LHEP, Albert Einstein Center, University of Bern, Bern, Switzerland}
\author{R.~deVoe}
 \affiliation{Physics Department, Stanford University, Stanford CA, USA}
\author{A.~Dobi}
 \affiliation{Physics Department, University of Maryland, College Park MD, USA}
\author{M.J.~Dolinski}
 \affiliation{Physics Department, Stanford University, Stanford CA, USA}
\author{A.~Dolgolenko}
 \affiliation{Institute for Theoretical and Experimental Physics, Moscow, Russia}
\author{M.~Dunford}
 \affiliation{Physics Department, Carleton University, Ottawa ON, Canada}
\author{W.~Fairbank Jr.}
 \affiliation{Physics Department, Colorado State University, Fort Collins CO, USA}
\author{J.~Farine}
 \affiliation{Department of Physics, Laurentian University, Sudbury ON, Canada}
\author{W.~Feldmeier}
\author{P.~Fierlinger}
 \affiliation{Technische Universit\"at M\"unchen, Physikdepartment and Excellence-Cluster ``Universe'', Garching, Germany}
\author{D.~Franco}
\author{G.~Giroux} 
\author{R.~Gornea}
 \affiliation{LHEP, Albert Einstein Center, University of Bern, Bern, Switzerland}
\author{K.~Graham}
 \affiliation{Physics Department, Carleton University, Ottawa ON, Canada}
\author{G.~Gratta}
 \affiliation{Physics Department, Stanford University, Stanford CA, USA}
\author{C.~Hall}
 \affiliation{Physics Department, University of Maryland, College Park MD, USA}
\author{K.~Hall}
 \affiliation{Physics Department, Colorado State University, Fort Collins CO, USA}
\author{C.~Hargrove}
 \affiliation{Physics Department, Carleton University, Ottawa ON, Canada}
\author{S.~Herrin}
 \affiliation{SLAC National Accelerator Laboratory, Stanford CA, USA}
\author{M.~Hughes}
  \affiliation{Department of Physics and Astronomy, University of Alabama, Tuscaloosa AL, USA}
\author{A.~Johnson}
 \affiliation{SLAC National Accelerator Laboratory, Stanford CA, USA}
\author{T.N.~Johnson}
 \affiliation{Physics Department and CEEM, Indiana University, Bloomington IN, USA}
\author{A.~Karelin}
 \affiliation{Institute for Theoretical and Experimental Physics, Moscow, Russia}
\author{L.J.~Kaufman}
 \affiliation{Physics Department and CEEM, Indiana University, Bloomington IN, USA}
\author{A.~Kuchenkov}
 \affiliation{Institute for Theoretical and Experimental Physics, Moscow, Russia}
\author{K.S.~Kumar}
 \affiliation{Physics Department, University of Massachusetts, Amherst MA, USA}
\author{D.S.~Leonard}
 \affiliation{Department of Physics, University of Seoul, Seoul, Korea}
\author{F.~Leonard}
 \affiliation{Physics Department, Carleton University, Ottawa ON, Canada}
\author{D.~Mackay}
 \altaffiliation{Now at KLA-Tencor, Milpitas CA, USA}
 \affiliation{SLAC National Accelerator Laboratory, Stanford CA, USA}
\author{R.~MacLellan}
 \affiliation{Department of Physics and Astronomy, University of Alabama, Tuscaloosa AL, USA}
\author{M.~Marino}
 \affiliation{Technische Universit\"at M\"unchen, Physikdepartment and Excellence-Cluster ``Universe'', Garching, Germany}
\author{B.~Mong}
 \affiliation{Department of Physics, Laurentian University, Sudbury ON, Canada}
\author{M.~Montero D\'{i}ez}
 \affiliation{Physics Department, Stanford University, Stanford CA, USA}
\author{A.R.~M\"{u}ller}
 \altaffiliation{Now at Kayser-Threde, Munich, Germany}
 \affiliation{Physics Department, Stanford University, Stanford CA, USA}
\author{R.~Neilson}
 \altaffiliation{Now at Department of Physics, University of Chicago, Chicago IL, USA}
 \affiliation{Physics Department, Stanford University, Stanford CA, USA}
\author{R.~Nelson}
 \affiliation{Waste Isolation Pilot Plant, Carlsbad NM, USA}
\author{A.~Odian}
 \affiliation{SLAC National Accelerator Laboratory, Stanford CA, USA}
\author{I.~Ostrovskiy}
\author{K.~O'Sullivan}
 \affiliation{Physics Department, Stanford University, Stanford CA, USA}
\author{C.~Ouellet}
 \affiliation{Physics Department, Carleton University, Ottawa ON, Canada}
\author{A.~Piepke}
 \affiliation{Department of Physics and Astronomy, University of Alabama, Tuscaloosa AL, USA}
\author{A.~Pocar}
 \affiliation{Physics Department, University of Massachusetts, Amherst MA, USA}
\author{C.Y.~Prescott}
 \affiliation{SLAC National Accelerator Laboratory, Stanford CA, USA}
\author{K.~Pushkin}
 \affiliation{Department of Physics and Astronomy, University of Alabama, Tuscaloosa AL, USA}
\author{P.C.~Rowson}
 \affiliation{SLAC National Accelerator Laboratory, Stanford CA, USA}
\author{J.J.~Russell}
 \affiliation{SLAC National Accelerator Laboratory, Stanford CA, USA} 
\author{A.~Sabourov}
 \affiliation{Physics Department, Stanford University, Stanford CA, USA}
\author{D.~Sinclair}
 \altaffiliation{Also TRIUMF, Vancouver BC, Canada}
 \affiliation{Physics Department, Carleton University, Ottawa ON, Canada}
\author{S.~Slutsky}
 \affiliation{Physics Department, University of Maryland, College Park MD, USA}
\author{V.~Stekhanov}
 \affiliation{Institute for Theoretical and Experimental Physics, Moscow, Russia}
 \author{T.~Tolba}
 \affiliation{LHEP, Albert Einstein Center, University of Bern, Bern, Switzerland}
\author{D.~Tosi}
 \affiliation{Physics Department, Stanford University, Stanford CA, USA}
\author{K.~Twelker}
 \affiliation{Physics Department, Stanford University, Stanford CA, USA}
\author{P.~Vogel}
 \affiliation{Kellogg Lab, Caltech, Pasadena, CA, USA}
\author{J.-L.~Vuilleumier}
 \affiliation{LHEP, Albert Einstein Center, University of Bern, Bern, Switzerland}
\author{A.~Waite}
 \affiliation{SLAC National Accelerator Laboratory, Stanford CA, USA}
\author{T.~Walton}
 \affiliation{Physics Department, Colorado State University, Fort Collins CO, USA}
\author{M.~Weber}
 \affiliation{LHEP, Albert Einstein Center, University of Bern, Bern, Switzerland}
\author{U.~Wichoski}
 \affiliation{Department of Physics, Laurentian University, Sudbury ON, Canada}
\author{J.~Wodin}
 \affiliation{SLAC National Accelerator Laboratory, Stanford CA, USA}
\author{J.D~Wright}
 \affiliation{Physics Department, University of Massachusetts, Amherst MA, USA}
\author{L.~Yang}
 \affiliation{Physics Department, University of Illinois, Urbana-Champaign IL, USA}
\author{Y.-R.~Yen}
 \affiliation{Physics Department, University of Maryland, College Park MD, USA}
\author{O.Ya.~Zeldovich}
 \affiliation{Institute for Theoretical and Experimental Physics, Moscow, Russia}

\collaboration{EXO Collaboration}

\date{\today}

\begin{abstract}

We report on a search for neutrinoless double-beta decay of $^{136}$Xe with EXO-200.   No signal is observed for an exposure of 32.5\,kg-yr, with a background of $\sim 1.5\times 10^{-3}$\,kg$^{-1}$yr$^{-1}$keV$^{-1}$ in the $\pm 1\sigma$ region of interest.  This sets a lower limit on the half-life of the neutrinoless double-beta decay $T_{1/2}^{0\nu\beta\beta}$($^{136}$Xe) $>$~1.6~$\times$~10$^{25}$\,yr (90\% C.L.), corresponding to effective Majorana masses of less than 140--380\,meV, depending on the matrix element calculation.

\end{abstract}


\maketitle

Double-beta decay is a rare nuclear decay observable in those even-even nuclides where $\beta$ decay is either energetically forbidden or highly spin suppressed. The hypothetical neutrinoless ($0\nu\beta\beta$) decay mode can only occur for massive Majorana neutrinos~\cite{schechter_valle}. This decay may be mediated by the exchange of a Majorana neutrino or by other new particles.  The recent discovery of neutrino mass in oscillation experiments~\cite{RPP} makes the search for the Majorana nature of neutrinos particularly relevant and timely.  The $0\nu\beta\beta$ decay rate is related to the square of an effective Majorana neutrino mass $\langle m \rangle_{\beta\beta}$ by the product of phase space and a nuclear matrix element squared.   Kinematic measurements restrict the neutrino mass scale to be below $\mathcal{O}(1\ {\rm eV})$~\cite{kine}, leading to $0\nu\beta\beta$ half lives beyond $10^{24}$\,yr. A search for $0\nu\beta\beta$ decay in $^{76}$Ge has claimed a positive observation~\cite{Klapdor} with 
$T^{0\nu\beta\beta}_{1/2}$($^{76}$Ge)$ = (2.23^{+0.44}_{-0.31})\times 10^{25}$\,yr, implying $\langle m \rangle_{\beta\beta} = 0.32\pm 0.03$\,eV for the nuclear matrix element given in \cite{QRPA2}.

The exceedingly long half lives of interest for $0\nu\beta\beta$ decay require large detectors using isotopically enriched sources, radio-clean construction techniques and the ability to actively reject remaining backgrounds.  The $0\nu\beta\beta$ decay results in a discrete electron sum energy distribution centered at the Q-value ($Q_{\beta\beta}$).  The allowed, yet also rare, two neutrino double beta ($2\nu\beta\beta$) decay is characterized by a continuous sum energy spectrum ending at $Q_{\beta\beta}$.   The $2\nu\beta\beta$ decay has been observed in many isotopes~\cite{RPP} and, recently, in $^{136}$Xe with the EXO-200 detector~\cite{EXO200_2nu}, later confirmed in~\cite{KLZ_2nu}.  The decay electron sum energy allows discrimination between the two modes.

EXO-200, described in detail in~\cite{EXO200_det1}, uses xenon both as source and detector for the two electrons emitted in its $\beta\beta$ decay. The detector is a cylindrical homogeneous time projection chamber (TPC)~\cite{TPC}.  It is filled with liquefied xenon ($^{enr}$LXe) enriched to (80.6$\pm$0.1)\% in the isotope $^{136}$Xe. The remaining 19.4\% is $^{134}$Xe, with other isotopes present only at low concentration. EXO-200 is designed to minimize radioactive backgrounds, maximize the $^{enr}$LXe fiducial volume, and provide good energy resolution at the $^{136}$Xe Q-value of $2457.83\pm0.37$ keV~\cite{Redshaw}. Energy depositions in the TPC produce both ionization and scintillation signals. The TPC configuration allows for three-dimensional topological and temporal reconstruction of individual energy depositions. This ability is essential for discriminating $\beta\beta$ decays from residual backgrounds dominated by $\gamma$s. 

The cylindrical TPC is divided into two symmetric volumes separated by a cathode grid. Each end of the TPC is instrumented with 38 charge induction (V) and 38 charge collection (U) wire triplets.  The U and V wire grids, crossing at $60^{\circ}$, provide stereoscopic information for charge depositions.  At each end of the TPC there are $\sim$250 Large Area Avalanche Photodiodes (LAAPDs) \cite{APDs} that record the 178\,nm scintillation light.  A drift field of 376\,V/cm is applied in the TPC volume.  All signals are digitized at 1\,MS/s.  All detector components were carefully selected to minimize internal radioactivity~\cite{activity}. The TPC is mounted in the center of a low-background cryostat.  At least 50\,cm (25\,cm) of HFE-7000 fluid~\cite{HFE} (lead) shield the TPC from external radioactivity.  Calibration sources can be inserted at various positions immediately outside of the TPC. The clean room module housing the TPC is surrounded on four sides by 50\,mm thick plastic-scintillator cosmic-ray veto panels, which are (95.5~$\pm$0.6)\% efficient. EXO-200 is located at a depth of $1585^{+11}_{-6}$ m.w.e. at the WIPP in New Mexico, USA. 

EXO-200 started taking low background data in late May 2011.  The data presented here were collected from September 22, 2011 to April 15, 2012, for a total of 2,896.6 hours live time under low background conditions.   
During the same period, 376.8 hours of calibration data were collected with 
three $\gamma$ sources at three positions.   Most of these data was taken with a $^{228}$Th source for the primary purpose of measuring the free electron lifetime ($\tau_{\rm e}$) in the TPC.  This is achieved by recording the charge collected for 2615\,keV full absorption events occurring at different z-positions.  The resulting $\tau_{\rm e}$ was $\sim$200~$\mu$s 
for the summer 2011 \cite{EXO200_2nu} but climbed to $\sim$3\,ms upon increasing the Xe recirculation flow for the current dataset.
  
\begin{figure}
\includegraphics[width=3.4in]{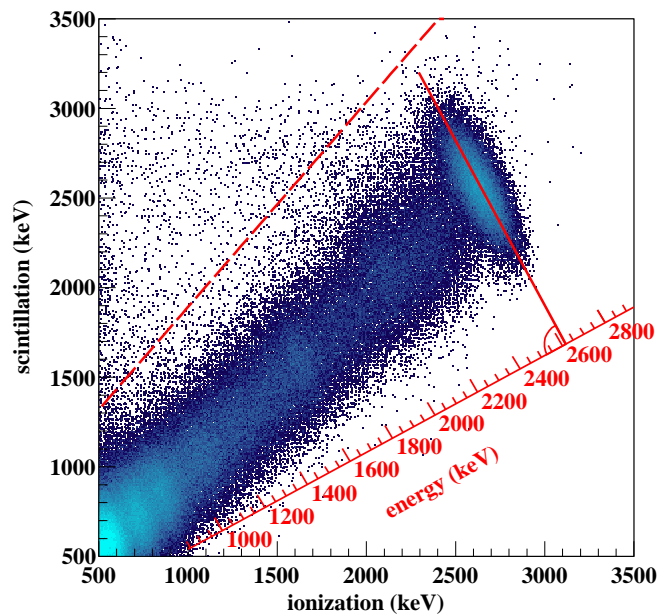}
\caption{Correlation between ionization and scintillation for SS events from a $^{228}$Th source. The energy resolution is considerably improved by forming the linear combination of both measurements.
Events in the top-left quadrant are due to incomplete charge collection and are rejected by the cut (dashed line), removing only 0.5\% of the total.}
\label{fig:correlation}
\end{figure}

Offline event reconstruction proceeds in three stages: signal finding, parameter estimation, and clustering. Charge signals on the U wires and scintillation signals on the two LAAPD planes are found using a matched filter technique. The filter yields time estimates for both ionization and scintillation channels.  The time information from the U wires is used to search for induction signals in V wire waveforms.    A signal un-shaping algorithm is applied to the U wire waveforms, optimizing the discrimination between single-site (SS) and multiple-site (MS) energy depositions. Candidate U wire signals are fit to template waveforms modeling the measured transfer functions so that the signal amplitudes can be extracted for energy estimation.   Amplitudes are corrected, channel-by-channel, for electronic gains determined from radioactive source calibration.   U wire and V wire signals are then combined into charge clusters using timing information from the fits, and associated with the nearest (earlier in time) summed scintillation signal.  Each cluster energy is corrected for position-dependent charge losses due to finite $^{enr}$LXe purity and for the shielding grid inefficiency of the V wire plane. This procedure, introducing the only time-dependent correction, yields a reconstructed ionization energy and three dimensional position information for each charge cluster.  An efficiency loss is incurred by events for which 3D reconstruction is not possible as these are rejected from this analysis.  

The LAAPD sum signal is corrected for variations of gain and spatial light collection efficiency within the TPC.  A trilinear interpolation of the 2615\,keV $\gamma$ SS full absorption peak scintillation signal, recorded for 1352 locations throughout the $^{enr}$LXe during $^{228}$Th calibration source deployments, is used to construct the correction function.

\begin{figure}
\includegraphics[width=3.4in]{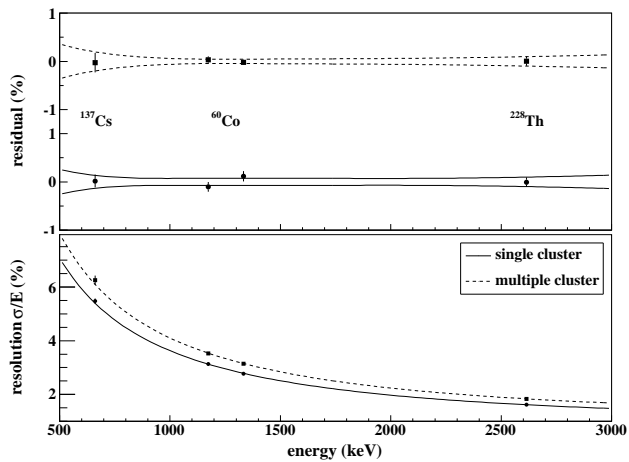}
\caption{Top: systematic uncertainty bands on the energy calibration residuals, using 
the full energy reconstruction described for the three $\gamma$ sources.  For both SS (solid) and MS (dashed) the position of the four $\gamma$ lines is consistent with the calibration model within $\leq$ 0.1\%.  Bottom: energy resolution for various sources along with a fit to the empirical model (see text).}
\label{fig:eres}
\end{figure}

Figure~\ref{fig:correlation} shows the energy of events as measured by the ionization and scintillation channels while the $^{228}$Th source was deployed.  As first discussed in~\cite{Conti_etal} and evident from the tilt of the 2615\,keV full absorption ellipse in the figure, the magnitude of the two signals is anticorrelated.  The 2D SS and MS energy spectra are independently rotated and projected onto a new (1D) energy variable in such a way as to minimize the width of the 2615\,keV $\gamma$ line.  Energy spectra from $^{137}$Cs, $^{60}$Co, and $^{228}$Th sources are produced using this method and then, the positions of the full absorption peaks at 662, 1173, 1333 and 2615\,keV are fit.  The energy calibration function that converts the rotated energy estimator into keV has a small quadratic term.  The residuals, defined as $(E_{\text{fit}} - E_{\text{true}})/E_{\text{true}}$, and the uncertainty bands are shown in Figure~\ref{fig:eres} (top).  The energy resolution, parameterized as  $\sigma^2 = a\sigma_e^2+bE+cE^2$, is plotted versus energy in Figure~\ref{fig:eres} (bottom).   Here $\sigma_e$ is the electronic noise contribution,  $bE$ represents statistical fluctuations in the ionization and scintillation, and $cE^2$ is a position- and time-dependent broadening.  It is found that $\sigma/E $~=~1.67\% (1.84\%) for SS (MS) spectra at $Q_{\beta\beta}$, dominated by the noise and broadening terms.  

\begin{figure}
\includegraphics[width=3.4in]{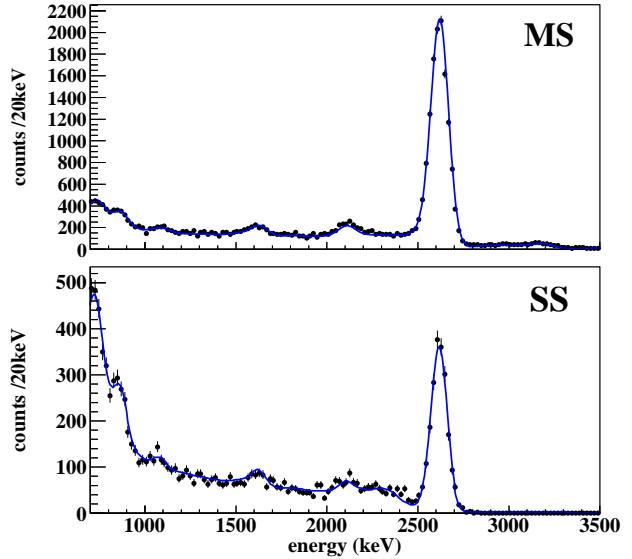}
\caption{MS (top) and SS (bottom) energy spectra from a $^{228}$Th calibration run.  The PDF generated from simulation (line) is fit to the data floating the normalization of the MS and SS spectra independently to illustrate the agreement with the rendered spectral shape.  The limited ability of the model to reproduce the absolute source rate is quantified in the text and is not illustrated by the figure.}
\label{fig:single-multi}
\end{figure}

The ability of the TPC to identify SS and MS interactions is used to separate $\beta$ and $\beta\beta$ decays in the bulk xenon from multiple site $\gamma$ interactions.  The clustering, currently applied in 2D, has a separation resolution of 18\,mm in the U-dimension and 6\,mm in z (drift time).  The event sharing between SS and MS energy spectra is demonstrated for a $^{228}$Th source in Figure \ref{fig:single-multi}.  The SS and MS spectra are compared to probability density functions (PDFs) generated by GEANT4~\cite{GEANT4} simulations (MC).  Simulated charge depositions are transported to the wires where signals are generated using a model of the readout electronics.  Noise is added and signals are reconstructed in the same manner as for data.  After events are designated as either SS or MS, the total collected energy is convolved with the energy resolution function shown in Figure \ref{fig:eres}.  This procedure reproduces the shape of the $^{60}$Co and $^{228}$Th source spectra (see Figure~\ref{fig:single-multi} for $^{228}$Th).  The MC reproduces the fraction of SS events, defined as $N_{\text{SS}}/(N_{\text{SS}}+N_{\text{MS}})$, to $\pm$8.5\%.    In addition, for $^{228}$Th ($0\nu\beta\beta$) events the simulation predicts a 70\% (71\%) efficiency for the requirement that events are fully reconstructed in 3D.  The simulation predicts the absolute, NIST-traceable, activity of all sources to within $\pm$9.4\%.  These variations are used as measures for the systematic uncertainties in the final fit.  The 71\% MC estimate for the $0\nu\beta\beta$ efficiency is further verified over a broad energy range by comparing the $2\nu\beta\beta$ MC efficiency with low-background data.  Although the $2\nu\beta\beta$ spectrum has vanishing statistical power at $Q_{\beta\beta}$, the efficiency is found to be a smooth function of the energy and agrees with the simulated efficiency to within the $\pm$9.4\% overall scale uncertainty mentioned above.

\begin{figure}
\includegraphics[width=3.4in]{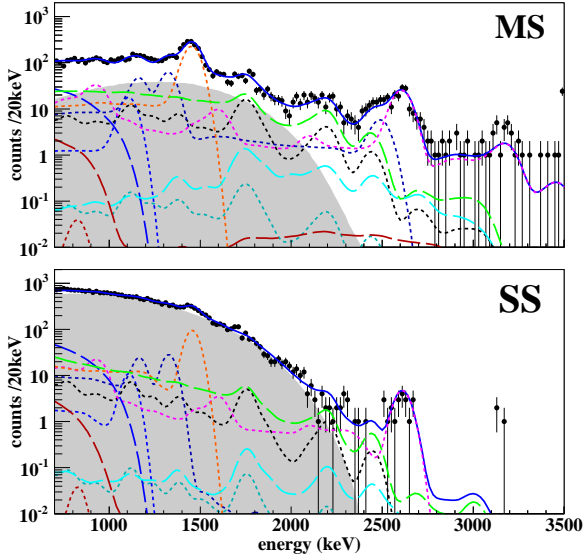}
\caption{MS (top) and SS (bottom) energy spectra.  The best fit line (solid blue) is shown.  The background components are $2\nu\beta\beta$ (grey region), $^{40}$K (dotted orange), $^{60}$Co (dotted dark blue), $^{222}$Rn in the cryostat-lead air-gap (long-dashed green), $^{238}$U in the TPC vessel (dotted black), $^{232}$Th in the TPC vessel (dotted magenta), $^{214}$Bi on the cathode (long-dashed cyan), $^{222}$Rn outside of the field cage (dotted dark cyan), $^{222}$Rn in active xenon (long-dashed brown), $^{135}$Xe (long-dashed blue) and $^{54}$Mn (dotted brown). The last bin on the right includes overflows (none in the SS spectrum).}
\label{fig:low-back}
\end{figure}

\begin{figure}
\includegraphics[width=3.4in]{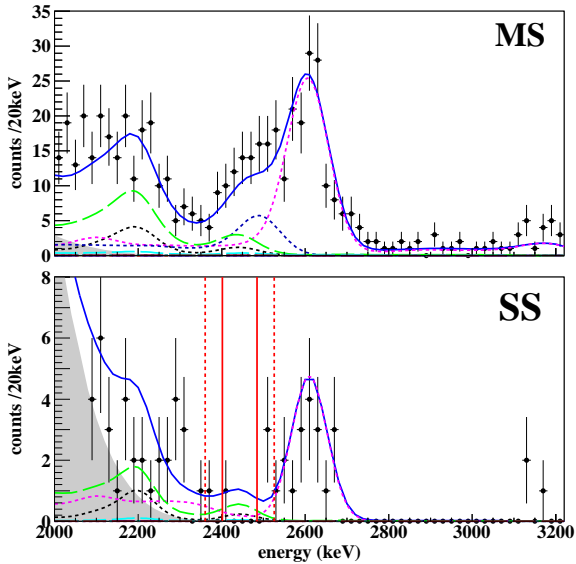}
\caption{Energy spectra in the $^{136}$Xe $Q_{\beta\beta}$ region for MS (top) and SS (bottom) events.  The 1 (2)$\sigma$ regions around $Q_{\beta\beta}$ are shown by solid (dashed) vertical lines.  The $0\nu\beta\beta$ PDF from the fit is not visible.  The fit results have the same meaning as in Figure~\ref{fig:low-back}.}
\label{fig:blip}
\end{figure}

The fiducial volume used in this analysis contains 79.4\,kg of $^{136}$Xe (3.52$\times$10$^{26}$ atoms), corresponding to 98.5\,kg of active $^{enr}$LXe.  The trigger is fully efficient above 700\,keV.  The cut represented by the dashed diagonal line in Figure~\ref{fig:correlation} eliminates a population of events due to interactions in the $^{enr}$LXe region for which the charge collection efficiency is low, leading to an anomalous light-to-charge ratio.  This cut also eliminates $\alpha$ decays from the low background data, but causes only a negligible loss of efficiency for $\gamma$- and $\beta$-like events.  Cosmic-ray induced backgrounds are removed using three time-based cuts. Events preceded by a veto hit within 25\,ms are removed (0.58\% dead time).  Events occurring within 60\,s after a muon track in the TPC are also eliminated (5.0\% dead time).  Finally, any two events that occur within 1\,s of each other are removed (3.3\% dead time).  The combination of all three cuts incurs a total dead time of 8.6\%.  The last cut, combined with the requirement that only one scintillation event per frame is observed, removes $\beta$-$\alpha$ decay coincidences due to the time correlated decay of the $^{222}$Rn daughters $^{214}$Bi and $^{214}$Po.  Alpha spectroscopic analysis finds $360 \pm 65 \;\mu$Bq of $^{222}$Rn in the $^{enr}$LXe, that is constant in time.

The SS and MS low background spectra are shown in Figure~\ref{fig:low-back}.  Primarily due to bremsstrahlung, a fraction of $\beta\beta$ events are MS.  The MC simulation predicts that 82.5\% of $0\nu\beta\beta$ events are SS.  Using a maximum likelihood estimator, the SS and MS spectra are simultaneously fit with PDFs of the $2\nu\beta\beta$ and $0\nu\beta\beta$ of $^{136}$Xe along with PDFs of various backgrounds.  Background models were developed for various components of the detector.  Results of the material screen campaign, conducted during construction, provide the normalization for the models.  The contributions of the various background components to the $0\nu\beta\beta$ and $2\nu\beta\beta$ signal regions were estimated using a previous generation of the detector simulation~\cite{EXO200_det1}.  For the reported exposure, components found to contribute $<0.2$ counts ($0\nu\beta\beta$) and $<50$ counts ($2\nu\beta\beta$), respectively, were not included in the fit.  For the current exposure, the background model treats the activity of the $^{222}$Rn in the air-gap between the cryostat and the lead shielding as a surrogate for all $^{238}$U-like activities external to the cryostat, because of their degenerate spectral shapes and/or small contributions.  A possible energy offset and the resolution of the $\gamma$-like spectra are parameters in the fit and are constrained by the results of the source calibrations.  The fraction of events that are classified as SS for each of the $\gamma$-like PDFs is constrained within $\pm$8.5\% of the value predicted by MC.  This uncertainty is set by the largest such deviation measured with the source calibration spectra.  The SS fractions for $\beta$- and $\beta\beta$-like events are also constrained in the fit to within $\pm$8.5\% of the MC predicted value.  As a cross-check, the constraint on the $2\nu\beta\beta$ SS fraction is released in a separate fit of the low background data.  The SS fraction is found to agree within 5.8\% of the value predicted by the MC simulation.

\begin{figure}
\includegraphics[width=3.4in]{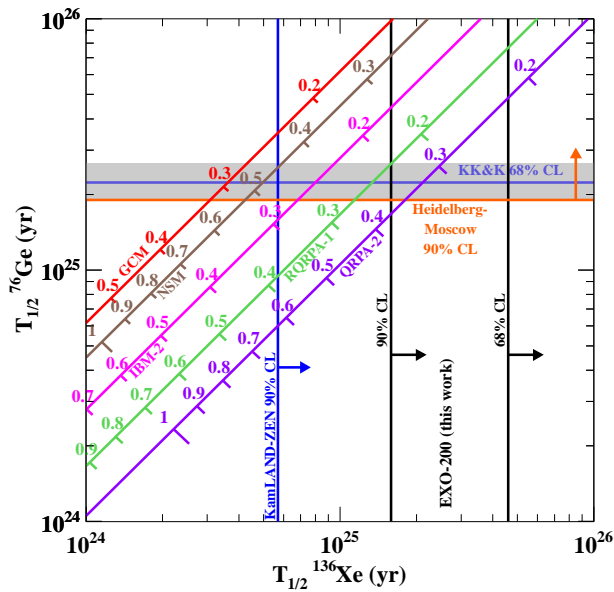}
\caption{Relation between the $T^{0\nu\beta\beta}_{1/2}$ in $^{76}$Ge and $^{136}$Xe for different matrix element calculations (GCM~\cite{GCM}, NSM~\cite{NSM}, IBM-2~\cite{IBM-2}, RQRPA-1~\cite{QRPA1} and QRPA-2 \cite{QRPA2}).   For each matrix element $\langle m \rangle_{\beta\beta}$ is also shown (eV).  The claim~\cite{Klapdor} is represented by the grey band, along with the best limit for $^{76}$Ge~\cite{heimo}.  The result reported here is shown along with that from~\cite{KLZ_2nu}.}
\label{fig:comparison}
\end{figure}

The $\beta\beta$ energy scale is a free parameter in the fit, so that it is constrained by the $2\nu\beta\beta$ spectrum.  The fit reports a scale factor of 0.995~$\pm$\,0.004.   The uncertainty is inflated to $\pm$\,0.006 as a result of an independent study of the possible energy scale differences between $\gamma$- and $\beta\beta$-like energy deposits.   The $2\nu\beta\beta$ PDF is produced using the Fermi function calculation given in~\cite{VogelFermiFunction}.  Tests using a slightly different spectral form~\cite{Iachello} were performed and found to contribute $<$0.001\% to the predicted location of $Q_{\beta\beta}$.  Finally, the stability of this energy scale correction was verified by repeating the fit with increasing thresholds up to 1200\,keV.

For the best-fit energy scale and resolution the $\pm 1\sigma$ and $\pm 2 \sigma$ regions around $Q_{\beta\beta}$ are shown in Figure~\ref{fig:blip}.   The number of events observed in the SS spectrum are 1 and 5, respectively, with the 5 events in the $\pm 2\sigma$ region accumulating at both edges of the interval.  Therefore, no evidence for $0\nu\beta\beta$ decay is found in the present data set.  The lower limit on $T^{0\nu\beta\beta}_{1/2}$ is obtained by the profile likelihood fit to the entire SS and MS spectra.  Systematic uncertainties are incorporated as constrained nuisance parameters.  The fit yields an estimate of $4.1\pm0.3$ background counts in the $\pm 1\sigma$ region, giving an expected background rate of $(1.5\pm0.1)\times 10^{-3}$\,kg$^{-1}$yr$^{-1}$keV$^{-1}$.  It also reports $0\nu\beta\beta$ decay limits of $<2.8$ counts at 90\% C.L. ($<1.1$ at 68\% C.L.).  This corresponds to a $T^{0\nu\beta\beta}_{1/2} > 1.6\times 10^{25}$\,yr at 90\% C.L. ($T^{0\nu\beta\beta}_{1/2} > 4.6\times 10^{25}$\,yr at 68\% C.L.).  Toy MC studies confirm the coverage of this method as suggested by \cite{Rolke}.  The same fit also reports $T_{1/2}^{2\nu\beta\beta} =(2.23\pm 0.017~{\rm stat.}\pm 0.22~{\rm sys.})\times 10 ^{21}$\,yr, in agreement with~\cite{EXO200_2nu} and \cite{KLZ_2nu}.   The levels of contamination from $\gamma$-emitting nuclides are found to be consistent with material screening estimates~\cite{activity}.  The addition to the fit of a PDF for $^{137}$Xe produces a 13\% higher limit on $T^{0\nu\beta\beta}_{1/2}$ at 90\% C.L..  In the absence of an independent constraint on this cosmogenic background, the smaller limit is reported.

The result from the likelihood fit is shown in Figure~\ref{fig:comparison}, along with the recent constraint for $^{136}$Xe~\cite{KLZ_2nu} and the best limit~\cite{heimo} and claimed detection~\cite{Klapdor} for $^{76}$Ge.   
The present result contradicts~\cite{Klapdor} at 68\% C.L. (90\% C.L.) for the nominal values of  all (most) matrix element calculations considered~\cite{GCM,NSM,IBM-2,QRPA1,QRPA2} and provides upper bounds to Majorana neutrino masses between 140 and 380\,meV at 90\% C.L..

\begin{acknowledgments}
EXO-200 is supported by DoE and NSF in the United States, NSERC in Canada, SNF in 
Switzerland, NRF in Korea and RFBR in Russia.   This research used resources of the National Energy Research Scientific Computing Center (NERSC). The collaboration gratefully acknowledges the WIPP for the hospitality and G.~Walther (Stanford) for discussions on statistical methods.
\end{acknowledgments}


\end{document}